\RequirePackage{fix-cm}
\documentclass{svjour3}
\smartqed  
\usepackage{graphicx}
\usepackage{dcolumn}
\usepackage{bm}
\usepackage{amsmath}
\usepackage{epstopdf}
\usepackage{subfig}
\usepackage{xcolor}%

\begin{document}
\title{A numerical characterisation of unconfined strength of weakly consolidated granular packs and its effect on fluid-driven fracture behaviour.}

\author{Paula A. Gago
\and
Charalampos Konstantinou
\and
Giovanna Biscontin
\and
Peter King}

\institute{Paula A. Gago \at
              Department of Earth Science and Engineering, Imperial College London, SW7 2AZ, UK.\\
              \email{p.gago@imperial.ac.uk} 
                        \and
                        Charalampos Konstantinou \at
                        Department of Engineering, University of Cambridge, CB2 1PZ, UK.\\
                        \email{ck494@cam.ac.uk}
\and
Giovanna Biscontin \at
                        Department of Engineering, University of Cambridge, CB2 1PZ, UK.\\
\and
          Peter King \at
          Department of Earth Science and Engineering, Imperial College London, SW7 2AZ, UK.\\
            }

\date{Received: date / Accepted: date}
\maketitle

\begin{abstract}

Soft or weakly-consolidated sand refers to porous materials composed of particles (or grains) weakly held together to form a solid but that can be easily broken when subjected to stress.
These materials do not behave as conventional brittle, linear elastic materials and the transition between these two regimes cannot usually be described using poro-elastic models.
Furthermore, conventional geotechnical sampling techniques often result in the destruction of the cementation and recovery of sufficient intact core is, therefore, difficult. 
This paper studies a numerical model that allows us to introduce weak consolidation in granular packs.
The model, based on the $LIGGGHTS$ open source project, simply adds an attractive contribution to particles in contact. 
This simple model allow us to reproduce key elements of the behaviour of the stress observed in compacted sands and clay, as well as in poorly consolidated sandstones. 
The paper finishes by inspecting the effect of different consolidation levels in fluid-driven fracture behaviour.
Numerical results are compared against experimental results on bio-cemented sandstones.

\end{abstract}

\section{Introduction}

Soft sands are granular materials in which particles are weakly held together resulting in very low stiffness.
Mechanisms that hold particles together include weak grain-to-grain cementation, external compressing forces and the physical interlocking arising from the irregular shape of the grains.  
Many engineering applications,  such as groundwater remediation, slope stabilisation, hydrocarbon extraction, or in situ leaching from mining, involve soft-sands and thus, the ability to predict their response to fluid injection is of great importance. 

These materials usually have low unconfined compressive strength and high permeability, although the literature lacks precise definitions for these ranges. 
They are intermediate between soils and rocks, sharing common characteristics with both \cite{Collins2009,Sitar1980}. 
Therefore, their behaviour is expected to be in-between sand and competent sandstone \cite{Nakagawa}.
An encompassing description for them is poorly-consolidated weakly-cemented sandstones:
They have low strength and stiffness, poor core integrity and stress dependent porosity and permeability. 

Experimental evidence has shown that fluid-driven fractures in soft sands are often shaped as flow channels (narrow paths presenting higher permeability than the matrix) with some degree of branching \cite{huang2012granular,johnsen2006pattern,Konstantinou2020b,chang2004hydraulic,Chang2003}. 
Elasto-plastic deformation likely plays a significant role in soft sand fracture initiation and propagation.
Although it is frequently assumed that shear mechanisms dominate the fracture tip dynamics, it is not clear whether tensile fracturing takes place. 
The mode of failure is particularly important as it is seen to describe the fracturing behaviour. In cohesionless sands, tensile stress cannot be transmitted and therefore failure shear occurs, however in soft rocks of low strength, dislocation of particles along with pure fractures might take place.
Some authors, identified this and linked the mode of failure with the unconfined compressive strength (UCS) of the host rock \cite{Olson2011}. It is therefore particularly important to test materials across various strengths in order to examine this transition in behaviour from pure shear failure to pure tensile failure.
The research on this topic remains very limited.
The majority of the aforementioned studies makes use of cohesionless sand  \cite{zhang2013coupled,johnsen2006pattern,huang2012granular,Chang2003,chang2004hydraulic,Pater2007,Dong2010,Dong2008,Xu2010,Konstantinou2020b} whose ``softness'' derives from inter-locking due to the particle shape, without cementation between the grains.

Contrary to the case of rock specimens, with virtually limitless quantities of a material for testing purposes, of any shape and size, obtaining soft rocks samples is particularly challenging: Conventional geotechnical sampling techniques often result in destruction of the cementation and recovery of sufficient intact core is, therefore, difficult. 
In addition, material properties are expected to change upon chemical and mechanical degradation.
As a consequence, good information about the stress-strain behaviour is rare \cite{Sitar1980}.
The task of generating artificial specimens that closely resemble the engineering properties of natural geological formations has been central in the fields of soil and rock mechanics \cite{Maccarini1987,Wygal1963,Vogler2017,Whiffin2004,Konstantinou2020b,Konstantinou2020}.

Konstantinou et al. \cite{Konstantinou2020}, used a bio-cementation technique termed as microbially induced carbonate precipitation ($MICP$) to generate artificially cemented rocks with mechanical and physical properties closely resembling those of natural weakly cemented carbonate sandstones. Those properties were well defined and reproducible. 
This technique involves the use of urease-producing bacterial strains which are flushed through the granular material together with a cementation solution supply consisting of urea and a calcium source. As a result, calcium carbonate precipitates bonding the particles \cite{Whiffin2004,DeJong2006}. The method allows for control of the cementation level and, therefore, materials of various strengths can be produced.
The preparation of such samples is, nevertheless, time consuming and weaker samples are harder to obtain because of practical limitations such as disaggregation at the grain scale during extraction from the molds. 

This paper uses a discrete element method ($DEM$) to model soft sands/poorly consolidated sandstones. 
The model, although simple, successfully reproduces key elements characteristic of fracture development in these materials.
Numerical results are compared against those obtained experimentally via $MICP$.

The structure of this work is as follows: 
Section \ref{sec:bio} presents the experimental $MICP$ technique and the main experimental observations relevant to the present study.
The numerical model is described in Sec. \ref{sec:model-and-setup}.
Section \ref{sec:stress-strain-test}, addresses the preparation of the ``numerical'' probes and the results for the stress-strain tests on them.
The results are discussed in the context of the information provided by the experimental evidence. 
To finish, preliminary results of the effect of the consolidation on the hydraulic fracture response of these materials (Sec. \ref{sec:fracture}) is analysed.

\section{Bio-cemented sandstones} \label{sec:bio}

The experimental samples used in this work are weakly cemented carbonate sandstones. 
They were created using microbially induced carbonate precipitation ($MICP$) following the experimental protocol presented in \cite{Konstantinou2020,Konstantinou2020b}.
The authors have conducted a parametric study to characterise the engineering properties of these bio-treated products across a range of cementation levels (from about $3\%$ to $11\%$  mass concentration of calcium carbonate). The main properties considered are unconfined compressive strength ($UCS$) and porosity. 

\begin{figure}
\centering
\subfloat[]{\includegraphics[width=1.\columnwidth]{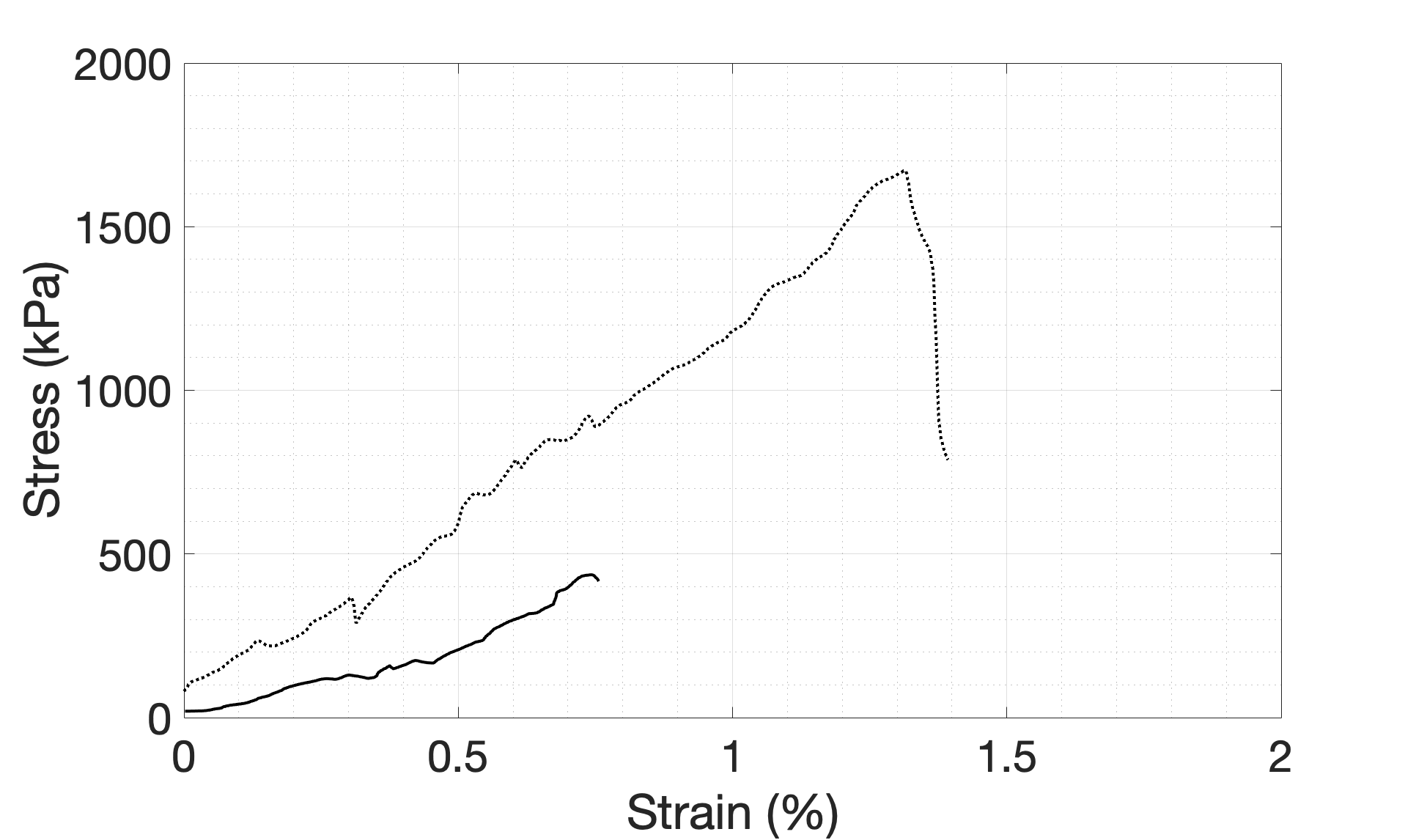}}\\
\subfloat[]{\includegraphics[width=0.4\columnwidth]{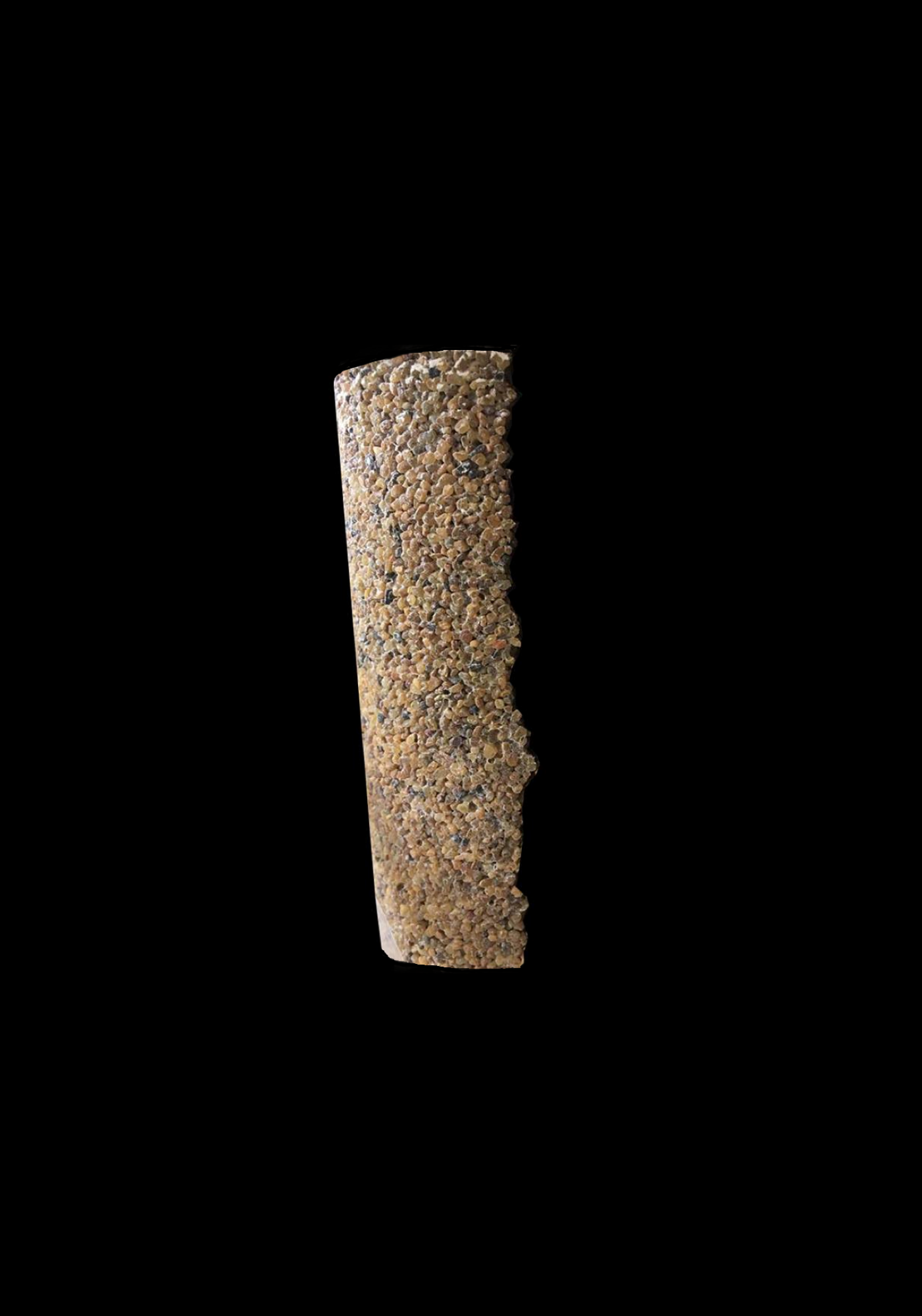}}
\subfloat[]{\includegraphics[width=0.4\columnwidth]{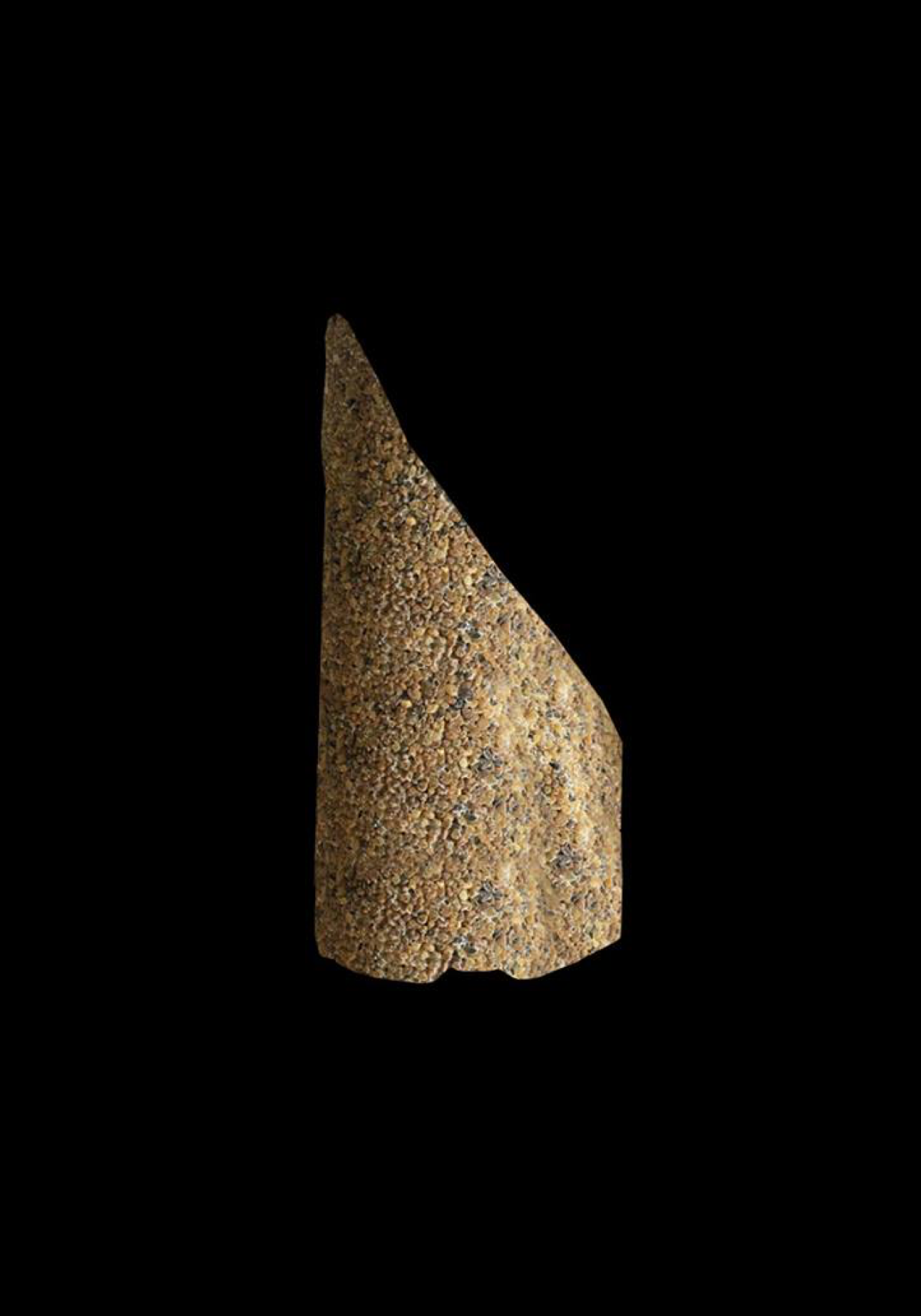}}
    \caption{(a) Examples of stress strain curves obtained in experimental $UCS$ testing for samples with low (5.5 \%) and high (10\%) cementation levels (continuous and dotted curves, respectively).
    (b-c) Characteristic failure  patterns for the specimens during the compression test: (b) tensile failure (left) and (c) shear failure (right).}
    \label{fig:strength}
\end{figure}

The $UCS$ tests were performed on oven-dried cylindrical specimens (diameter of $70~$mm, height $140~$mm) across various cementation levels.
The base material used for these experiments was a uniformly graded coarse sand with an average particle size  of $0.180~$mm ($D_{50}$), for which there is very limited data in the literature as most studies use fine graded sands.
The samples were prepared and tested according to \cite{Measurements2004}. 
The loading was displacement-controlled with a rate of $1.14~$mm/min. 
Examples of stress strain curves obtained for low and high cementation levels are shown in Fig.  \ref{fig:strength} (a). 

Figure \ref{fig:strength} (b-c) shows the characteristic modes of failures obtained during unconfined compression tests. At very low cementations (up to 3 \%,  corresponding to UCSs of less than 200 kPa) the specimens disaggregated at the grain scale during the compression test because they were very weak. For cementation levels of 3\% to 8\%, corresponding to UCS values of $200$ to $1000~$kPa, the specimens showed a characteristic axial splitting mode.
At degrees of cementation above 8\%, shearing along a single plain was evident.

\begin{figure}
\centering
\includegraphics[width=1\columnwidth]{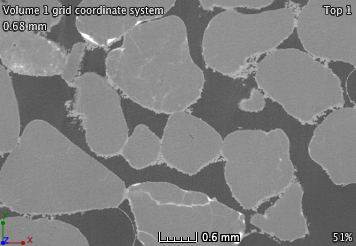}
    \caption{MicroCT imaging of a moderately cemented (5 \%) bio-treated coarse sand specimen: the carbonate crystals accumulate on the surface of the grains whilst it acts as a bridge for transmission of stresses at the contact points creating the sense of overlapping. X-ray $\mu$-CT high energy micro-tomography scanner (X-Tec Systems) settings: resolution at 8 um, 160 kV, 110 uA with a use of a 0.5 mm copper filter.}
    \label{fig:CT}
\end{figure}

Figure \ref{fig:CT} shows a microCT image for an $MICP$ sample.
The distribution of the cementation within the granular network and the location of the carbonate crystals on the grain surfaces can be observed.
 The enhancement of strength is attributed to the amount of cementation that is present at the contact points between particles. This is because the strength is transmitted to the particles via those contact points, and any addition of carbonate crystals results in a higher strength at the macro-scale level. 
 The carbonate crystals themselves appear to be of the same size, much smaller compared to the size of the grains; however, they accumulate as the amount of cementation increases generating clusters. A detailed discussion on this is presented by \cite{Konstantinou2020,Konstantinou2020b}. 

\begin{figure}
\centering
\includegraphics[width=1.\columnwidth]{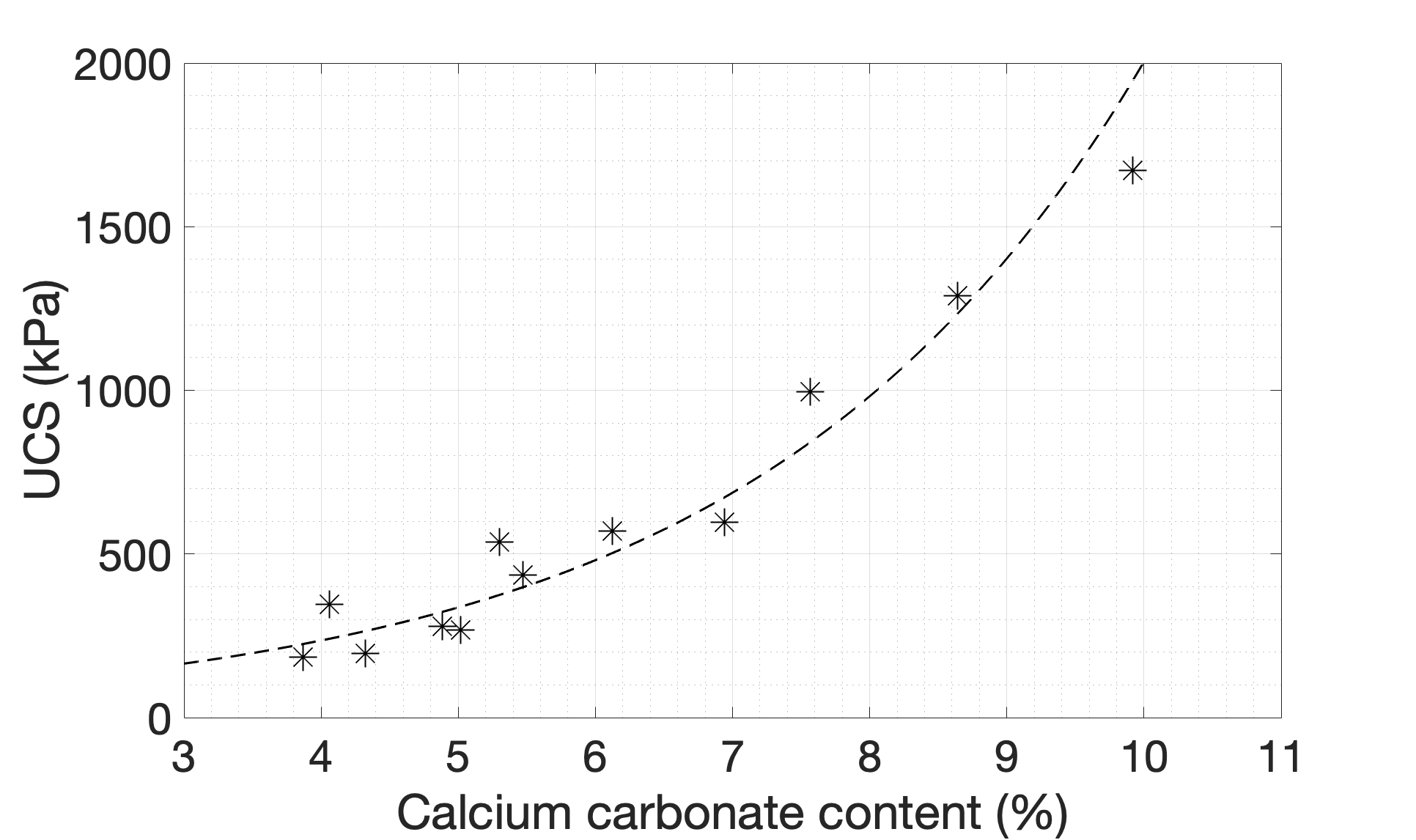}
    \caption{Strength of bio-cemented coarse sands as a function of the calcium carbonate content in the sample preparation. An exponential curve provided the best fit for this data: $UCS=56.544*exp(0.3568*C)$ where $C$ is the cementation level.}
    \label{fig:strength2}
\end{figure}

Figure \ref{fig:strength2} shows the behaviour of the unconfined compressive strength ($UCS$) of the bio-cemented coarse sands as a function of the calcium carbonate in the sample \cite{Konstantinou2020b,Konstantinou2020}.

The $UCS$ varies from about $100~$kPa to $1600~$kPa, which is a similar range to that found in natural weakly cemented sandstones reported in the literature.
The lower boundary of this range is, nevertheless, still uncertain, as weaker samples are harder to obtain, because of practical limitations.

\section{Numerical model}  \label {sec:model-and-setup}

We use the molecular dynamics of soft-spheres, discrete element method ($DEM$) model, provided by the $LIGGGHTS$ \cite{goniva2012influence} open source project. 
It implements a \textit{Hertz} potential \cite{brilliantov1996model} to model a ``repulsive'' force (due to volume exclusion) between two particles in contact and a simplified Johnson-Kendall-Roberts ($JKR$) \cite{johnson1971surface,johnson1996continuum} to model particle-particle cohesion (adhesion).
The normal component $F_{normal}$ of the repulsive force has a restoring term (that depends on $\delta^{3/2}$, with $\delta$ the linear overlap between the particles) and a dissipative term (proportional to the normal component of the relative velocity). 
The tangential component $F_{tang}$ also implements an elastic shear force, with a ``history'' effect that accounts for the tangential displacement between the particles during the time they are in contact, and a damping/viscous force, proportional to the tangential component of the relative velocity. 
The tangential (frictional) force is limited using $F_{tang} \leq \mu \times F_{normal}$, where $\mu$ is the friction coefficient.

The ``cohesive'' component, consists of adding an attractive normal force $F_{coh}=C_{h} \times A$ to hold together two particles in contact. 
Here $A$ is the cross sectional area of  the overlap between the particles and $C_{h}$ (in units of Pa) is called the cohesion energy density. 
This will be the control parameter for varying the system strength throughout this paper.

For a static sample (particle velocities equal to zero) the normal force for two particles $ij$ in contact can then be written as: 
\begin{equation}
 f_{n}^{ij}=-\frac {4} {3} \big( \frac {2(1-\nu^2)^{2}}{Y} \big)^{-1} \big( \frac {1}{r_{i}}- \frac{1}{r_{j}} \big)^{-1/2} \ \delta^{3/2} + C_{h} A_{ij} 
 \label {eq:normal_forces}
\end{equation}
where $\delta$ is the linear overlap between the two particles in contact, $Y$($=1 \times 10^{7}~$Nm$^{-2}$) is the Young\textquoteright  s modulus, $\nu$ ($=0.15$) is the Poisson\textquoteright s ratio.
The model also uses a friction coefficient $\mu=0.7$ and a restitution coefficient $\epsilon = 0.055$.
The low value selected for the restitution coefficient helps to reach a static configuration sooner (by dissipating energy faster), but it does not affect the system dynamics.
(Details of the model implementation can be found at \cite{SKRL}.)

It is possible to see that this $f_{n}$ can reach positive (cohesion dominant) and negative (compression dominant) values, and that there will be an inter-play between the ``stress'' and cohesion in the system (and that will be discussed in Sec. \ref{sec:fracture}).

\section{Stress-strain test}  \label {sec:stress-strain-test}
 
The grains are spheres with diameters distributed as a truncated Gaussian distribution ($d=[0.9-1.8]~$mm with mean diameter $d=1.363~$mm and a standard deviation $\sigma_{d}\sim 0.3~$mm.)

\begin{figure}
\centering
\includegraphics[width=1.\columnwidth]{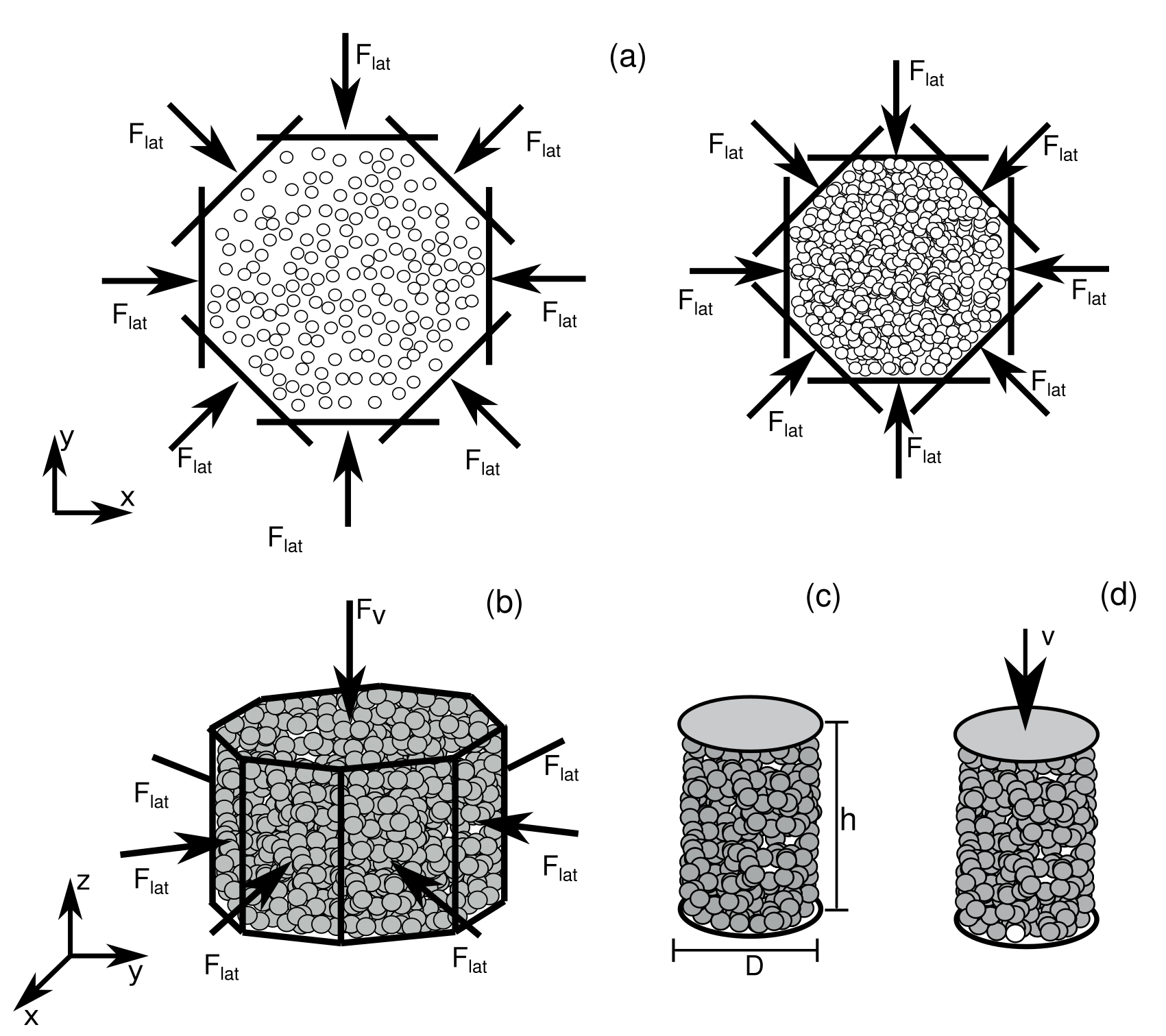}
    \caption{Schematic representation of the setup used to create the probes for the  stress-strain test. (a) and initially loose configuration of cohesionless grains, confined between two parallel, horizontal plates (in the $z$ direction) is laterally compressed ($F_{lat}$) until reaching a static, mechanically stable, dense condition. 
    (b) Cementation is added by setting $C_{h}>0$ (Eq. (\ref{eq:normal_forces})) while maintaining the effect of $F_{lat}$ and applying a vertical force $F_{v}=30~$N. All lateral walls are removed after a static configuration is achieved.
    (c) A cylinder of diameter $D=5~$cm and height $h-10~$cm is cut from the sample created in (b) (elimination the particles outside the region).
    (d) A constant displacement ($v=2.5~$mm s$^{-1}$) is imposed in the upper wall. Force acting on this wall is measured during the test.}
    \label{fig:numerical_setup}
\end{figure}

Cylinders of $h=10$cm height and $D=0.5~$cm diameter (formed by $N \approx 10^{5}$ grains) were constructed to perform the stress-strain tests.
The protocol followed to create these probes is presented in Fig. \ref{fig:numerical_setup}.

Initially, a loose configuration of $N=5 10^{5}$ spherical, non-cohesive ($C_{h}=0$) grains, is placed between two parallel plates separated by a constant distance $h$($=10~$cm) in the vertical ($z$) direction.
These grains are laterally compressed (in the $x-y$ plane) using $8$ piston-like walls applying a constant lateral force $F_{lat}$ until a mechanically stable configuration is reached (Fig. \ref{fig:numerical_setup} (a)).
This methodology was chosen to reduce stress anisotropy and minimise the creation of preferential stress directions in the system.
Once a mechanically stable configuration is reached, ``cementation'' is incorporated to the system by setting a value to the parameter $C_{h}>0$.
During this step the piston-like walls keep applying $F_{lat}$ and the upper wall applies a constant vertical force $F_{v}$($=30~$N) (Fig. \ref{fig:numerical_setup} (b)).
The system is allowed to relax under this condition until a new equilibrium state is achieved. 
All lateral walls are removed at this stage and a cylindrical core of $5~$cm diameter is obtained by ``deleting'' particles outside that region (Fig. \ref{fig:numerical_setup} (c)).

The system is again allowed to relax before the stress-strain test starts.
For the test, the upper horizontal wall moves at a constant speed $v$ ($=2.5~$mms$^{-1}$) (Fig. \ref{fig:numerical_setup} (d)), inducing deformation to the probe. 
The force acting on the wall due to the grains is measured and plotted against the strain $\epsilon=h-vt/h)$ (with $t$ being the time in seconds). 

Each complete test takes, depending on the cementation level of the sample ($C_{h}$), between one to three days (running on $32$ processors on Imperial College Research Computing Service (see DOI: 10.14469/hpc/2232)).
It is worth noting that this limitation in computational time, $v$ was chosen about ten times faster than that recommended in the literature \cite{Measurements2004} and results may be slightly affected by this.
In order to achieve lower strength levels, consistent with subsurface situations but hard to achieve experimentally, gravity was not considered, preventing that the sample breaks under its own weight.

\begin{figure}
\centering
\includegraphics[width=1.\columnwidth]{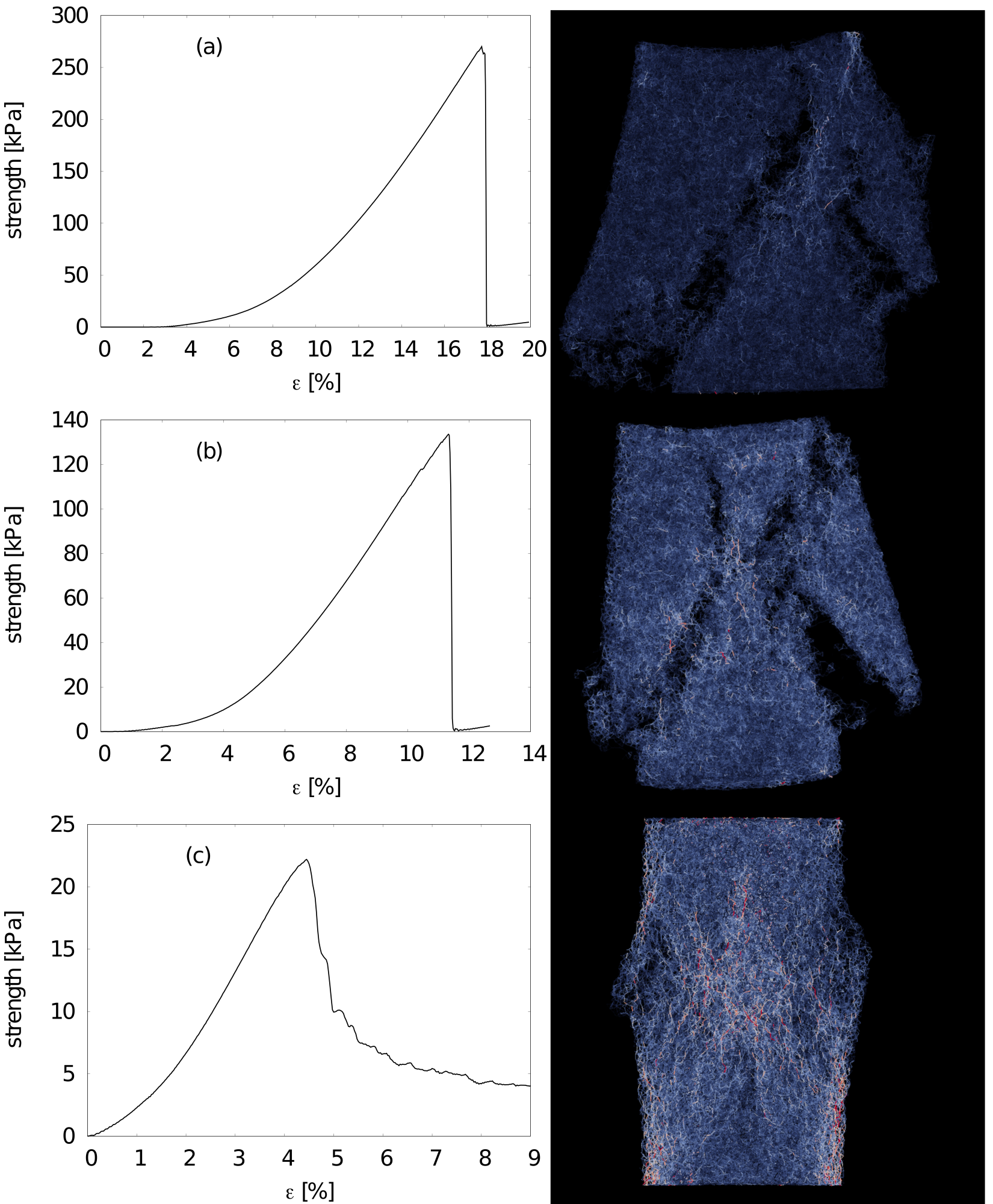}
    \caption{Left: stress-strain curves corresponding to different cohesion levels (from (a) to (e) $CH=5 \times 10^{5}$, $CH=4 \times 10^{5}$ $CH=3 \times 10^{5}$, $CH=2 \times 10^{5}$ $CH=1 \times 10^{5}$. Right: corresponding fractured probes (at the failure time). }
    \label{fig:UCS}
\end{figure}

Figure \ref{fig:UCS} shows, in the left column, the characteristic stress-strain curves for three different degrees of cementation (increasing from bottom to top). 
The same figure shows, in the  right column, the fracture behaviour for the corresponding probes.  

Key elements observed experimentally, are presented in these results.
One such behaviour is the effect known as post-peak strain softening.
It can be observed in  Fig. \ref{fig:UCS} (c) (and to a smaller extent in Fig. \ref{fig:UCS} (b)), as a soft decay in the stress supported for the system after the maximum strength value is achieved. 
This behaviour has been reported as characteristic of cemented soils and clay\cite{riman2014predicting,abdulla1997behavior} and attributed to the re-orientation of the sliding surfaces creating a residual shear strength (the lowest possible) \cite{culshaw2017material}.

It can also be noticed that a transition occurs as the material strength increases. This continuum deformation gives place to a more sudden ``brittle-like'' fracturing process (Fig. \ref{fig:UCS} (a)).
This type of behaviour can be compared with that displayed in Fig. \ref{fig:strength} (c), corresponding to moderate cementation levels in the $MICP$ samples ($\sim 300~$kPa), where an axial splitting mode was observed. 
At lower strengths or equivalently lower cementation levels, disaggregation at the grain scale was observed in the UCS experiments. This behaviour is mapped to the behaviour of the numerical setup where, expansion of the cylinder (consistent with Reynolds\textquoteright  dilatancy \cite{reynolds1885lvii}) is observed near the failure point. 
The focus on lower strengths or cementation levels, which were not possible in the experimental part because specimens of such low cohesion break when extracting from the mould, is one of the objectives of this study.

To assess the impact of the initial condition on the resulting strength of the sample, three different values were investigated for the confining force $F_{lat}$.

\begin{figure}
\centering
    \includegraphics[width=1.\columnwidth]{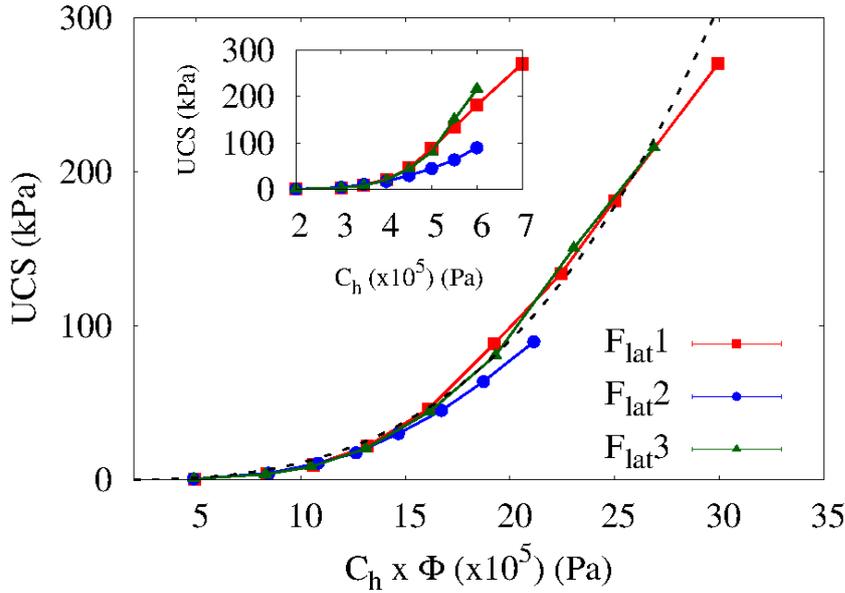}
    \caption{(a) Strength of the numerical samples as a function of the model cohesion energy density $C_{h}$ times $\Phi_{c}$, the density of contacts in the volume $\Phi_{c}=\phi \times z_{c}$ with $\phi$ the packing fraction and $z_{c}$ the mean coordination number of the particles in the system. 
    The different curves were obtained on samples built using different values of $F_{lat}1=33$N,$F_{lat}1=166$N,$F_{lat}1=333$N (Fig. \ref{fig:numerical_setup} (a)).
    The dashed black line corresponds to the function $UCS=0.01 \times (C_{h} \times \Phi)^{3}$ (obtained from fitting the data).
    Inset: Same than main figure but as a function $C_{h}$.}
    \label{fig:strengthC}
\end{figure}

The inset of Fig. \ref{fig:strengthC} shows $UCS$ for these different initial conditions as a function of the model parameter $C_{h}$ (Eq. (\ref{eq:normal_forces})).

In order to scale these results as a function of a more meaningful parameter, we define the density of contacts in the system  $\Phi_{c}$:
\begin{equation}
 \Phi_{c}=z_{c} \times \phi
\end{equation}
with $z_{c}$ the mean coordination number of the system and $\phi$ the corresponding packing fraction ($\phi=1-porosity$).

\begin{figure}
\centering
    \includegraphics[width=0.8\columnwidth]{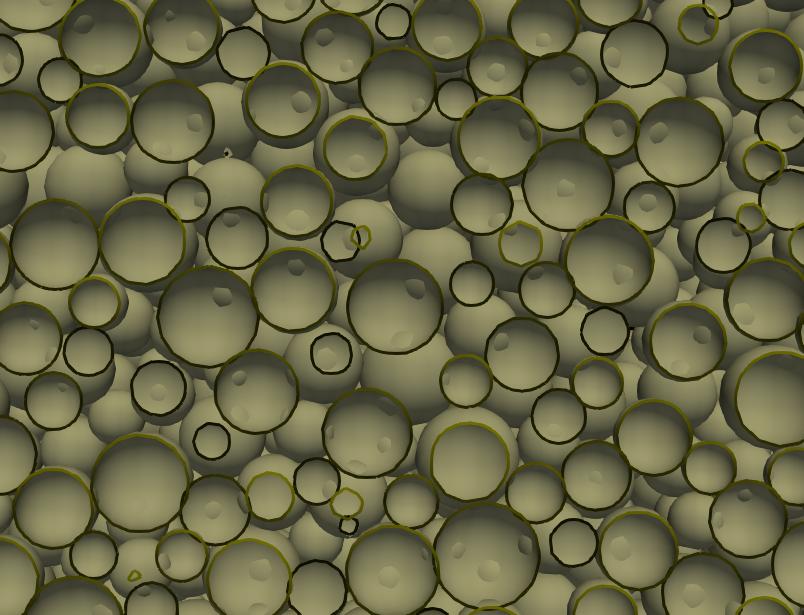}
    \caption{(a) Slice showing the ``microscopic structure'' of a numerical sample.}
    \label{fig:numerical_micro}
\end{figure}

Figure  \ref{fig:strengthC} shows $UCS$ as a function of $C_{h} \times \Phi_{c}$ for the same systems than in the inset. 
This parameter clearly gives a good agreement.
Figure \ref{fig:numerical_micro} contains a snapshot of an slice of the numerical probes.
Here, the overlap between particles gives a cohesive force as playing a similar role to the calcium carbonated bonds in $MICP$ samples (Fig. \ref{fig:CT}).

Figure \ref{fig:strengthC} shows that the maximum strength range reached by the numerical samples crosses the lower experimental end.
Therefore, the good agreement displayed between the fracture behaviour of on Figures \ref{fig:strength2} (a) and \ref{fig:UCS} (a) is encouraging of the adequacy of the model presented.

\section{Hydraulic fractures} \label{sec:fracture}

In this section, we use the cohesion model presented above to study the  behaviour of fluid-driven fractures created in quasi-two-dimensional Hele-Shaw cells.

Initial configurations are created following the protocol presented in section \ref{sec:model-and-setup}, but compressing $N=10^{4}$ loose grains between two parallel plates having a constant separation $h=1.5~$mm (in the vertical $z$ direction).
The effective cell diameter is $r \sim 4.8~$cm, (the cell is in fact octagonal). 
A central inlet of radius $r_{0}=3.5~$mm, is created during the compaction process by preventing particles from occupying the cylindrical region in the centre of the cell (using a wall-like constraint).
After reaching a static configuration (fig. \ref{fig:numerical_setup} (c)) all the walls in the system were fixed in position.
The volume of the cell is not allowed to change during fluid injection. 

Fluid injection is simulated using the $CFDEM$ \cite{kloss2012models,gago2020spatially} open source software.
This software couples $LIGGGTHS$ with $OPENFoam$, computational fluid dynamics ($CFD$) model to simulate spatially resolved solid-fluid interaction.
We simulate an incompressible fluid with viscosity $\nu=0.1~$cP, density $\rho=1000~$km$^{-3}$.
As this model needs to resolve the movement of the fluid around the solid interface a fine $CFD$ mesh is required (of at least four grid-blocks per particle diameter \cite{kloss2012models}). 
A mesh refinement of $1/6$ of the mean particle diameter was chosen. 

\begin{figure}
\centering
\includegraphics[width=1.\columnwidth]{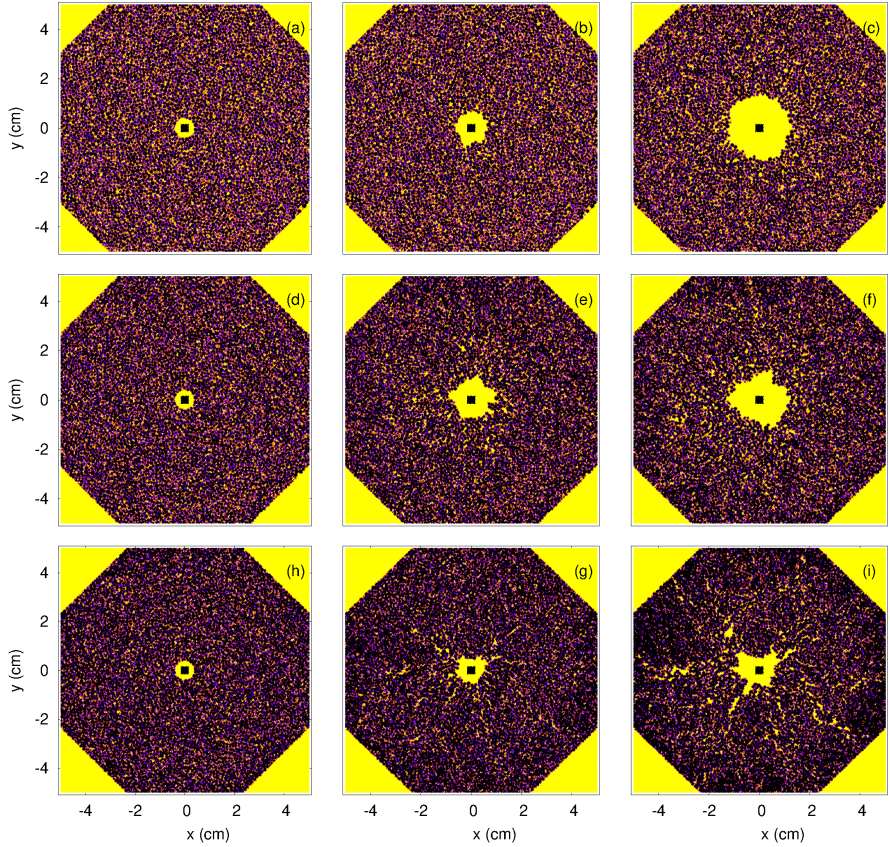}
    \caption{Fracture patterns obtained numerically. Injection pressures increases from left to right and consolidation degree from top to bottom. }
    \label{fig:numerical_fractures}
\end{figure}

Fluid is injected from the central inlet at a constant pressure  $p_{i}$ while the outlet (the external cell perimeter) is considered to be open to the atmosphere ($p_{outlet}=0$).
Figure \ref{fig:numerical_fractures} shows characteristic patterns of the fracture behaviour for  different injection pressures (increasing from left to right), and consolidation levels (increasing from top to bottom). 

These patterns show that the fracture patterns corresponding to more consolidated systems (bottom) tend to present more branched-like behaviour than those of unconsolidated ones (top).
This finding is not surprising: weaker materials spend more of the input energy in permanent deformation, while stiffer materials open up defects, those defects (or voids) increases the local pressure and ``self-drive'' the fracture development. 
This kind of behaviour is consistent with experimental observations.

\begin{figure}
\centering
\includegraphics[width=0.3\columnwidth]{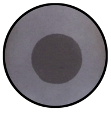}
\includegraphics[width=0.3\columnwidth]{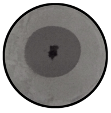}
\includegraphics[width=0.3\columnwidth]{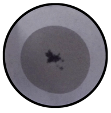}
\includegraphics[width=0.3\columnwidth]{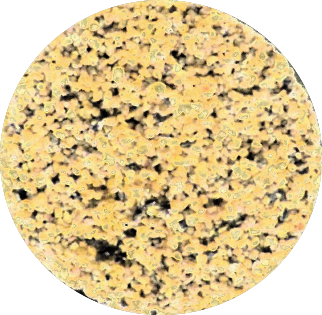}
\includegraphics[width=0.3\columnwidth]{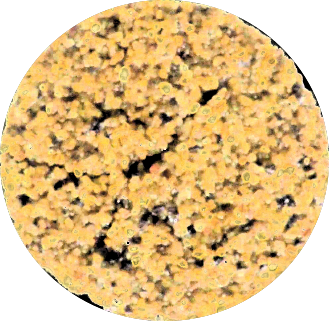}
    \caption{Top: An example of fracture behaviour in cohesionless fine sand, for three different values of the fluid pressure (increasing from left to right).
    Bottom: An example of an experimental injection test in a bio-cemented coarse carbonate coarse sandstone with strength of approximately $300-350$kPa.}
    \label{fig:experimental_fractures}
\end{figure}

Figure \ref{fig:experimental_fractures} (top) shows the fracture patterns obtained experimentally for an unconsolidated system with fine sand for three different values of the injection pressure (increasing from left to right).
The system was constrained by the boundary walls and the injection was from the bottom to the top. The borehole inlet was 1 cm away from the observation window (placed at the top). 
For the test with the lower applied inlet pressure, the fluid infiltrates through the permeable granular matrix without any visible fractures being induced. For the other two tests, where the inlet pressure was higher, particle rearrangement (grain dislocation) is observed resulting in an opening or borehole expansion.
As in Fig. \ref{fig:numerical_fractures} (top and middle), the fracture area is ``symmetrically'' distributed around the fluid inlet. Although the sand is cohesionless, it possesses some strength due to interlocking of the particles.
Figure \ref{fig:experimental_fractures} (bottom) shows a test before and after  a fracture was obtained on bio-cemented coarse sands with a $UCS \sim 300-350$ kPa. This experimental setup differs from the one used in unconsolidated sand as no walls were used to constrain the specimen (the specimen possesses strength that is higher than any potential reaction from the walls).
Multiple crack-like fractures can be observed (bottom right) without any evidence of grain dislocation or borehole expansion. The damage was extensive and was produced in a very small time window.
These patterns are similar to those in Fig. \ref{fig:numerical_fractures} (bottom).

In order to assess the effect of consolidation level in the development of the hydraulic fractures, a series of numerical experiments were performed.
A combination of confining stress $\Sigma$ (obtained by varying $F_{lat}$ in Fig. \ref{fig:numerical_setup} (a)) and material $UCS$ (obtained with different values of $C_{h}$) were considered. 
Material stress $\Sigma$ is defined as in \cite{gago2020stress} as the average value of the trace of the compressive stress tensor  $S^{i}_{ab}$ of the particles in the system: 
\begin{equation}
 S^{i}_{ab}=-0.5*(V^{i})^{-1}\sum_{c} r^{ic}_{a} f^{ic}_{b}
 \label{eq:stress}
\end{equation}
where  $i$ is the particle index and the sum runs over all the particles $c$ in contact with the particle $i$,  $V^{i}$ is the particle volume, $r^{ic}_{a}$ ($a=\{x,y,z\}$) is the component in the $a$ direction of the vector connecting the centre of the particle $i$ with the  centre of the particle $c$ and $f^{ic}_{b}$ is the $b$ component of the contact force.
Only positive contact forces (repulsive net force in Eq. (\ref{eq:normal_forces})) are considered in this definition.

\begin{figure}
\centering
\includegraphics[width=1.\columnwidth]{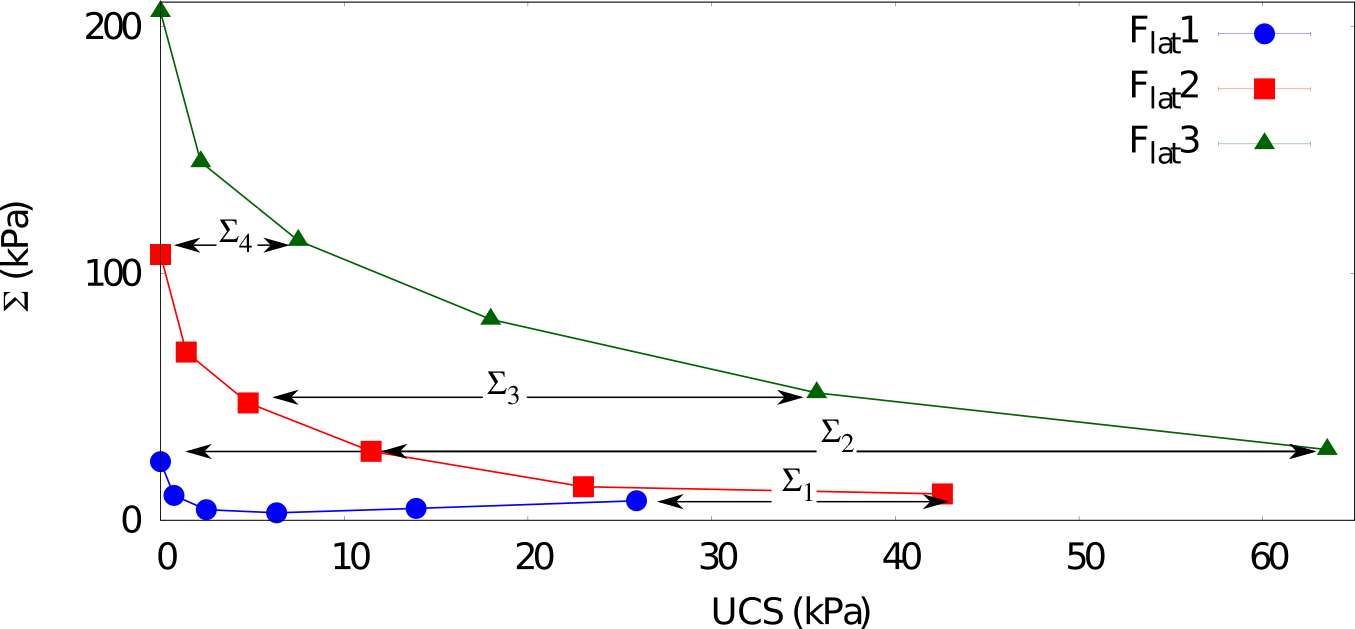}
    \caption{Compressive stress $\Sigma$ acting on the system as a function of the consolidation level (measured through the relation proposed in Fig. \ref{fig:strengthC}. As expressed in Eq. (\ref{eq:normal_forces}) compressive (repulsive) and cohesive (attractive) forces acting on the grains are not independent and initial conditions created under the same protocol ($F_{lat}$ do not necessarily present the same effective stress.}
    \label{fig:parametric}
\end{figure}

As Eq. (\ref{eq:normal_forces}) shows there is an interplay between  stress and strength values in this model: 
The addition of a ``cohesive'' component decreases the ``compressive'' component. 
Fig. \ref{fig:parametric} demonstrates this interplay. 
The value of $UCS$ was obtained through the power-law relation presented in Fig. \ref{fig:strengthC}.

\begin{figure}
\centering
\includegraphics[width=1.\columnwidth]{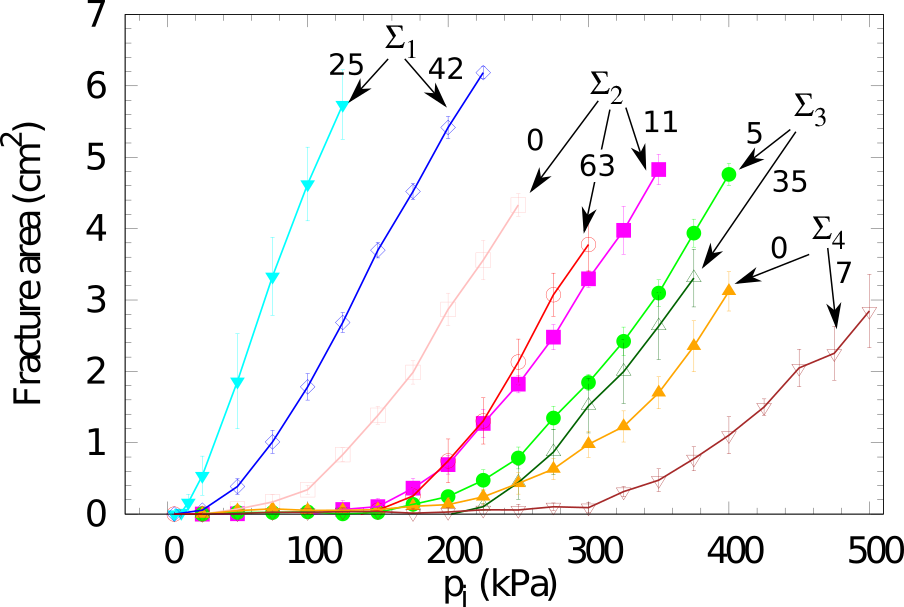}
    \caption{Fracture area for the numerical experiments, as a function of the inlet pressure for samples with different $UCS$ (numbers on the arrows, reported in kPa) and same stress $\Sigma$ (represented by the horizontal lines in Fig. \ref{fig:parametric}).}
    \label{fig:fracture}
\end{figure}

Fig. \ref{fig:fracture} shows the fracture area for the numerical experiments, as a function of the inlet pressure for samples with different cohesion level-s and the same compressive stress $\Sigma$ (represented by the horizontal lines in Fig. \ref{fig:parametric}).

Fracture area was measured using the protocol explained in Ref. \cite{gago2020fluid}. 
In short, the particle positions were mapped onto a $(500,500,8)$ cubic grid and each grid-block was set with a value in the range $[0-1]$ proportional to the amount of it covered by a particle.
Only the slice $z=4$ of the grid was considered (this is represented in Fig. \ref{fig:numerical_fractures}). 
Grid-blocks containing a fraction of particle smaller than $0.2$ were taken as fluid. 
To ignore the pore size volumes, only clusters of such grid-blocks, bigger than $10$ units were considered as a ``fracture''.
The fracture area as a function of $p_{i}$ was defined as the difference between the amount of those grid-blocks before and after the fluid injection.

This figure shows that, as expected, fractures in more consolidated materials require higher pressure to initiate, although, as has been discussed in \cite{gago2020stress}, the effective stress also strongly influenced fracture development. 
Curves corresponding to lower $\Sigma_{1}$ show a clear deviation for a change of consolidation of $\sim 25~$kPa, while for intermediate stress/cohesion this deviation is less evident ($\Sigma_{2}$ $UCS=63-11~$kPa and $\Sigma_{3}$ $UCS=5-35~$kPa)).
This deviation becomes evident again for higher stress and $UCS$ jumping from $0$ to a small, non-zero value.
A more systematic analysis of this dependence needs to be carried out.

\section{Conclusions}  \label{sec:conclusions}

We presented a discrete element method that allows for modelling key elements of the behaviour under stress of soft sands.
We have compared the numerical results with experimental results obtained on bio-cemented sandstones.
Although the range of strength for the numerical specimens are on the ``weaker'' end of the experimental samples, both overlap for moderate strength, allowing a proper validation of the model.
The model is successful in reproducing shear-failure, observed experimentally for materials with moderate strength ($UCS \sim 300$kPa) and characteristic branched-like fracture patterns driven by fluid-injection.
The model expands the experimentally possible strength values to an even lower range which could not be achieved in the experiments.
In this lower range, the model is able to account for the post-peak strain softening, characteristic of compacted sands.
Since it is believed that any transition in the hydraulic fracturing behaviour will occur in this lower region, the results presented here are encouraging that the method presented can help towards a better understanding of this transition.
A more systematic analysis of the fracture behaviour with consolidation/stress dependence, with special emphasis on the soft-brittle transitional $UCS$ values  will be addressed in future work.

\begin{acknowledgements}
The authors would like to acknowledge the funding and technical support from bp through the bp International Centre for Advanced Materials (bp-ICAM) which made this research possible.
Simulations were performed at the Imperial College Research Computing Service (see DOI: 10.14469/hpc/2232).
\end{acknowledgements}

\medskip
\bibliographystyle{unsrt}
\bibliography{./cites.bib}

\end{document}